\documentclass[apj]{emulateapj}

\usepackage{color}
\usepackage{amsmath}
\usepackage{natbib}
\usepackage{makeidx, xcolor}
\definecolor{linkc}{rgb}{0,0,1.0}
\usepackage[breaklinks, colorlinks, urlcolor={linkc}, citecolor={linkc}, linkcolor={linkc}]{hyperref}
\usepackage{etoolbox}

\makeatletter
\patchcmd{\NAT@citex}
  {\@citea\NAT@hyper@{%
     \NAT@nmfmt{\NAT@nm}%
     \hyper@natlinkbreak{\NAT@aysep\NAT@spacechar}{\@citeb\@extra@b@citeb}%
     \NAT@date}}
  {\@citea\NAT@nmfmt{\NAT@nm}%
   \NAT@aysep\NAT@spacechar\NAT@hyper@{\NAT@date}}{}{}

\patchcmd{\NAT@citex}
  {\@citea\NAT@hyper@{%
     \NAT@nmfmt{\NAT@nm}%
     \hyper@natlinkbreak{\NAT@spacechar\NAT@@open\if*#1*\else#1\NAT@spacechar\fi}%
       {\@citeb\@extra@b@citeb}%
     \NAT@date}}
  {\@citea\NAT@nmfmt{\NAT@nm}%
   \NAT@spacechar\NAT@@open\if*#1*\else#1\NAT@spacechar\fi\NAT@hyper@{\NAT@date}}
  {}{}

\makeatother

\newcommand{\BD}[1]{{\color{black} #1}}

\newcommand{\swift}{\emph{Swift}}
\newcommand{\rxte}{\emph{RXTE}}

\newcommand{\fermi}{\emph{Fermi}}
\newcommand{\suzaku}{\emph{Suzaku}}

\newcommand{\integral}{\emph{INTEGRAL}}
\newcommand{\xb}{{4U 0114+650}}

\begin{document}

\title{Evolution of Spin, Orbital, and Superorbital Modulations of 4U 0114+650}
\author{Chin-Ping Hu$^1$, Yi Chou$^2$, C.-Y. Ng$^1$, Lupin Chun-Che Lin$^3$, David Chien-Chang Yen$^4$}

\affiliation{$^1$ Department of Physics, The University of Hong Kong, Pokfulam Road, Hong Kong; {\color{blue} cphu@hku.hk}\\
$^2$ Graduate Institute of Astronomy, National Central University, Jhongli 32001, Taiwan \\
$^3$ Institute of Astronomy and Astrophysics, Academia Sinica, Taiwan \\
$^4$ Department of Mathematics, Fu Jen Catholic University, New Taipei City 24205, Taiwan \\
}
\submitted{ApJ, in press} 



\begin{abstract}
We report a systematic analysis of the spin, orbital, and superorbital modulations of \xb, a high-mass X-ray binary that consists of one of the slowest spinning neutron stars.  Using the dynamic power spectrum, we found that the spin period varied dramatically and is anticorrelated with the long-term X-ray flux variation that can be observed using the \emph{Rossi X-ray Timing Explorer} ASM, \swift\ BAT, and the Monitor of All-sky X-ray Image.  The spin-up rate over the entire dataset is consistent with previously reported values; however, the local spin-up rate is considerably higher. The corresponding local spin-up timescale is comparable to the \BD{local spin-up rate of OAO~1657$-$415}, indicating that \xb\ could also have a transient disk. Moreover, the spin period evolution shows two $\sim$1000-day spin-down/random-walk epochs that appeared together with depressions of the superorbital modulation amplitude. \BD{This implies that the superorbital modulation was closely related to the presence of the accretion disk, which is not favored in the spin-down/random-walk epochs because the accretion is dominated by the direct wind accretion.} The orbital period is stable during the entire time span; however, the orbital profile significantly changes with time. We found that the depth of the dip near the inferior conjunction of the companion is highly variable, which disfavors the eclipsing scenario.  Moreover, the dip was less obvious during the spin-down/random-walk epochs, indicating its correlation with the accretion disk. Further monitoring in both X-ray and optical bands could reveal the establishment of the accretion disk in this system. 
\end{abstract}

\keywords{X-rays:binaries --- stars: neutron --- accretion, accretion disks --- pulsars: individual: 4U 0114+650}

\section{Introduction}
Accretion-powered pulsars, usually found in X-ray binary systems, are powered by the gravitational potential energy of the accreting materials from the companion stars \citep{DavidsonO1973, LambPP1973}.  The spin period evolution of an accretion-powered pulsar highly depends on the accretion form, that is, wind-fed accretion or disk-fed accretion, as well as the strength of the neutron star magnetic dipole moment. The investigation of the spin period evolution in accreting pulsars is an important approach to understanding the effect of magnetic field and accreting materials on neutron stars. To date, more than 30 accreting hard X-ray pulsars have been regularly monitored with the Burst and Transient Source Experiment (BATSE) on board the \emph{Compton Gamma Ray Observatory} (\emph{CGRO}) and the Gamma-ray Burst Monitor (GBM) on board the \emph{Fermi Gamma-ray Space Telescope} for more than 10 years \citep{BildstenCC1997, FingerBN2009}. The comprehensive studies on these data provide us a detailed insight of the long-term behavior of accreting pulsars.

The source \xb, which was discovered in 1977 by the SAS-3 Galactic survey \citep{DowerKM1977}, is a high-mass X-ray binary (HMXB) system. It was originally recognized as a Be-star X-ray binary \citep{KoenigsbergerSS1983} and then revised to a supergiant X-ray binary \citep{CramptonHC1985, ReigCC1996}.  The compact object in this system is one of the slowest rotating pulsars.  A weak detection of a spin period of 895\,s was first claimed \citep{KoenigsbergerSS1983}.  Then, a period of $2.78$\,hr was found and further confirmed as the true spin period \citep{FinleyBC1992, HallFC2000}, although other interpretations such as tidally induced pulsations \citep{KoenigsbergerGM2006} have also been proposed.  The spin-up rate was first determined by \citet{HallFC2000} and then updated by many subsequent observations \citep[see, e.g.,][]{BonningF2005, FarrellSO2008, Wang2011}. \BD{Table \ref{spin_point_obs} lists previously detected spin frequencies of \xb\ by the \emph{Rossi X-ray Timing Explorer} (\rxte) PCA, the \integral\ IBIS, and the \suzaku\ XIS.} The spin period is not monitored by BATSE and GBM. However, it can be detected by all-sky monitoring programs such as the RXTE all-sky monitor (ASM) \citep{CorbetFP1999}. \citet{WenLC2006} systematically searched for periodicities in the ASM data with a time span of $\sim$ 8.5 years, and found that the pulsations of \xb\ were only detected in the first 4.5\,years of the ASM data. Dividing the ASM light curve with 0.5-year intervals, the spin period was found to change dramatically on a considerably short timescale \citep{WenLC2006}. The long-term evolution of the pulse period suggested that the spin-up rate is increasing, and the best-fit spin-up rate during 2003--2008 was $(1.09\pm0.13)\times10^{-6}$ s s$^{-1}$ \citep{Wang2011}.

\begin{deluxetable}{cccc}
\centering
\tiny
\tablecaption{Previous measurements of the spin period of \xb. \label{spin_point_obs}} 

\tablehead{\colhead{Time} & \colhead{Spin Per-} & \colhead{} &\colhead{} \\
\colhead{(MJD)} & \colhead{iod (s)} & iod (s) & \colhead{Reference} } 
\startdata
50371 & 9828(36) & \rxte\ PCA & \citet{HallFC2000} \\ 
52986 & 9612(20) & \integral\ IBIS & \citet{Wang2011} \\
53043 & 9600(20) & \integral\ IBIS & \citet{Wang2011} \\
53347 & 9605(14) & \integral\ IBIS & \citet{BonningF2005} \\
53354 & 9570(20) & \integral\ IBIS & \citet{Wang2011} \\
53520 & 9540(36) & \rxte\ PCA & \citet{FarrellSO2008} \\
53555 & 9555(15) & \integral\ IBIS & \citet{Wang2011} \\
53725 & 9520(20) & \integral\ IBIS & \citet{Wang2011} \\
53732 & 9518(11) & \rxte\ PCA & \citet{FarrellSO2008} \\
54580$^a$ & 9475(25) & \integral\ IBIS & \citet{Wang2011} \\
55763 & 9391.19 & \suzaku\ XIS & \citet{PradhanPP2015}
\enddata
\tablenotetext{a}{The time of this observation listed in Table 1 of \citet{Wang2011} is inconsistent with that in Figure 6 of the same paper. Therefore, we did not include it in our figure.}
\end{deluxetable}

The orbital period of this system determined from the radial velocity measurement of the companion LS I $+$65$^{\circ}$010 is $11.5983\pm0.0006$\,days with an orbital eccentricity $e=0.18\pm0.05$ \citep{CramptonHC1985,KoenigsbergerGM2006}. The X-ray orbital variability was first detected with \rxte\ ASM \citep{CorbetFP1999} and then confirmed as $11.599\pm0.005$\,days with an 8.5-year time baseline \citep{WenLC2006}. Although some of the previously reported values have a significant deviation, no clear orbital period change was ever reported. The X-ray orbital profile shows a sawtooth modulation that indicates a variable absorption by the stellar wind \citep{HallFC2000, GrundstromBG2007}, with a stable dip interpreted as an eclipse by the companion \citep{CorbetFP1999}.  However, \citet{FarrellSO2008} argued that the dip feature was not caused by the eclipse, and proposed an alternative scenario in which the neutron star passes through a heavily absorbing region close to the base of the stellar wind, which would explain the spectral variability.  Based on the difference between the orbital profile obtained with \rxte\ ASM and the \swift\ Burst Alert Telescope (BAT), as well as on the variability of the spectral index and the equivalent width of iron lines from a \suzaku\ observation, \citet{PradhanPP2015} also questioned the eclipsing scenario. They suggested that the sawtooth modulation was caused by the variation in mass accretion rate, and the dip at the inferior conjunction of the companion was caused by an increasing column density of the stellar wind. 

In addition to the spin and orbital periods, \xb\ also exhibits a long-term modulation with a period of 30.7\,days \citep{FarrellSO2006}. A change in the modulation profile was proposed to explain the null detection of the superorbital modulation in \citet{CorbetFP1999}.  \citet{KotzeC2012} showed that the superorbital modulation period was stable, with alternating strong and weak detections in the dynamic power spectrum. If the superorbital modulation of \xb\ was caused by the precession of a warped accretion disk as in Her X-1, SMC X-1, and LMC X-4, then a stable period is expected according to the prediction of a radiation-driven warp \citep{OgilvieD2001}.  On the other hand, \citet{MasettiOF2006} proposed that the superorbital modulation of \xb\ was probably produced by the precession of the neutron star spin axis. If this is true, there should be a connection between the superorbital modulation and the spin period although this has not yet been observed \citep{FarrellSO2006}. \citet{FarrellSO2008} suggested that the modulation may be associated with a superorbital phase-dependent Roche-lobe overflow, which was caused by a third donor star with an eccentric orbit. Hence, the origin of the superorbital modulation is still unclear. 

Our research uses all-sky monitoring data, including \rxte\ ASM and \swift\ BAT, to investigate the variability of modulation periods and X-ray profiles for the spin, orbital, and superorbital modulations in \xb.  We briefly introduce the observations and data of the two all-sky monitoring programs in Section \ref{observation}. The data analysis and main results, including the evolution of the spin period, the orbital period and profile, and the superorbital modulation, are described in Section \ref{analysis}.  We then discuss our discoveries in Section \ref{discussion}, including the connection between these three modulations and the origin of the orbital modulation.  Finally, we present a summary of this research in Section \ref{summary}.

\section{Observations}\label{observation}
The \rxte\ ASM was built for monitoring the variable and transient X-ray sources in the sky \citep{LevineBC1996}.  It consists of three cameras with proportional counters, and each of the cameras has a field of view of $6^{\rm{\circ}}\times90^{\rm{\circ}}$. The total collecting area is 90\,cm$^2$. It swept the entire sky every $\sim90$ minutes. The \rxte\ ASM has collected data during its mission life time from 1996 to early 2012. The energy range of the ASM is 1.5--12 keV, which can be further divided into three energy bands: 1.5--3, 3--5, and 5--12\,keV.  The dwell (each exposure) and one-day binned light curves for all monitored sources were archived and maintained by the Massachusetts Institute of Technology.  \BD{To eliminate the possible contaminations from the bad data points, we filtered these data points with uncertainties that are 3$\sigma$ higher than the mean uncertainty.} We analyzed the dwell light curve because the spin period of \xb\ can be resolved with this timing resolution. The gain of the ASM changed significantly after 2010 \citep{LevineBC2011}. For this reason, we excluded the data collected after MJD 55200.

The BAT is an instrument of the \swift\ satellite that triggers alerts of gamma-ray bursts and has monitored known X-ray sources since 2004 \citep{BarthelmyBC2005}.  The energy range of the BAT is 15--150 keV, which is considerably higher than that of the ASM. It is a coded mask imaging telescope with a field of view of 1.4 steradian, an angular resolution of $\sim$20\arcmin, and a large collecting area of 5200\,cm$^2$. \swift\ is a low-Earth-orbit satellite that has the ability to scan the entire sky every $\sim96$ minutes. The hard X-ray transient monitor program \citep{KrimmHC2013} provided ``daily'' and ``orbital'' light curves for monitored sources in the energy range of 15--50\,keV.  We analyzed the orbital light curve in this study \BD{and applied the same criterion as for the ASM data to filter the light curve}. \xb\ was monitored with the BAT with a total exposure time of $\sim$37\,Ms, and the exposure time of each orbit was from $64$ to $\sim2800$\,s. 

The Monitor of All-sky X-ray Image (MAXI), which is mounted on the Japanese Experimental Module of the International Space Station, is designed for monitoring the variability of the X-ray sources in the sky \citep{MatsuokaKU2009}. It consists of two slit cameras: the Solid-state Slit Camera with a collecting area of 200\,cm$^2$ in the energy range of 0.5--12\,keV, and the Gas Slit Camera with a collecting area of 5350\,cm$^2$ in the energy range of 2--30\,keV. The archival standard product containing one-day binned and one-orbit binned light curves are maintained by RIKEN, JAXA, and the MAXI team. Because the spin period of \xb\ cannot be significantly detected by MAXI, we only use the MAXI data in the phase evolution analysis of the orbital period (see Section \ref{evo_orbital}).

\section{Data Analysis and Result}\label{analysis}

\subsection{Evolution of the Spin Period}\label{evo_spin}


\begin{figure*}[t]
\begin{minipage}{0.47\linewidth}
\includegraphics[width=0.95\textwidth]{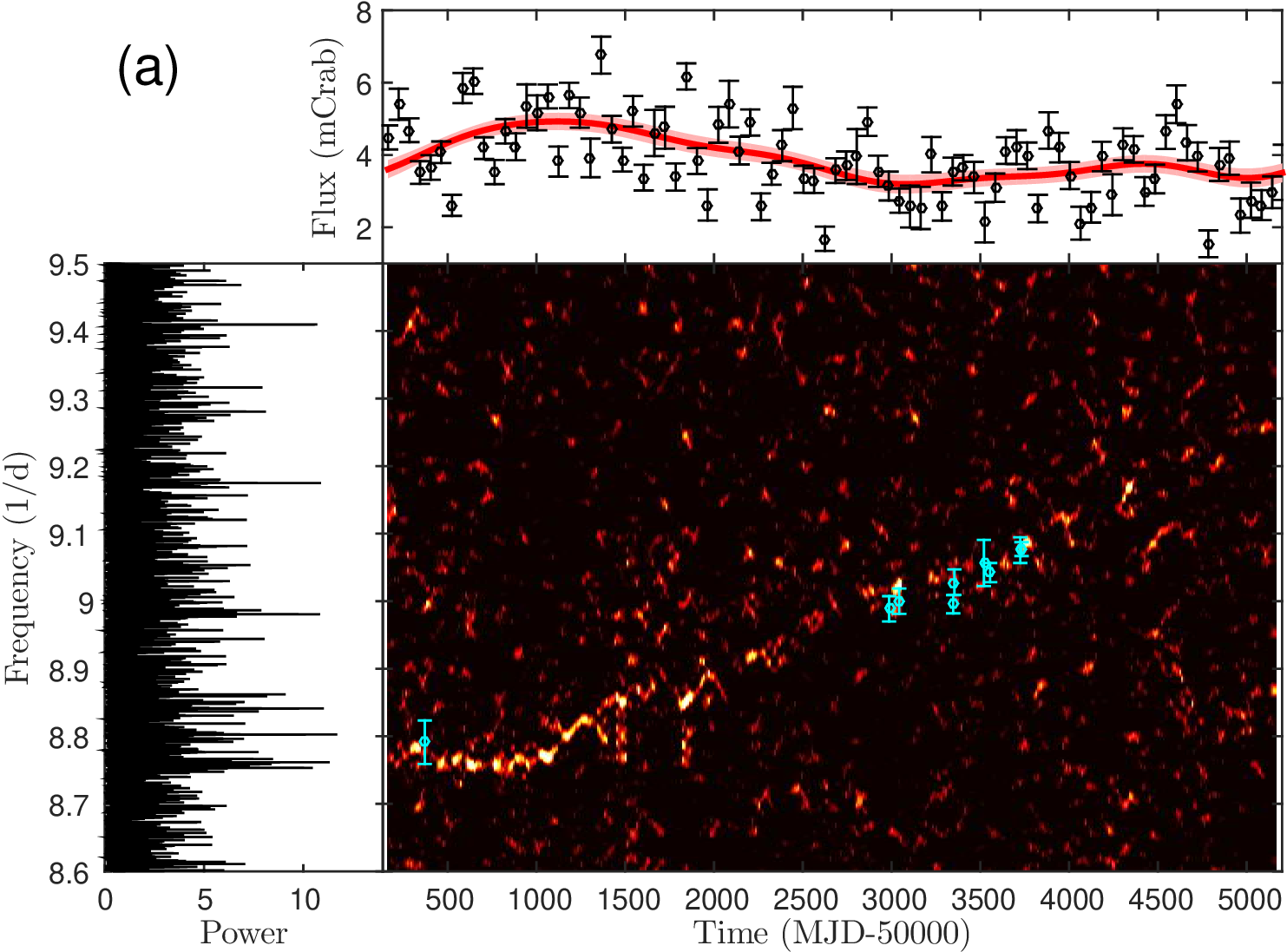}
\hspace{0.5cm}
\includegraphics[width=0.95\textwidth]{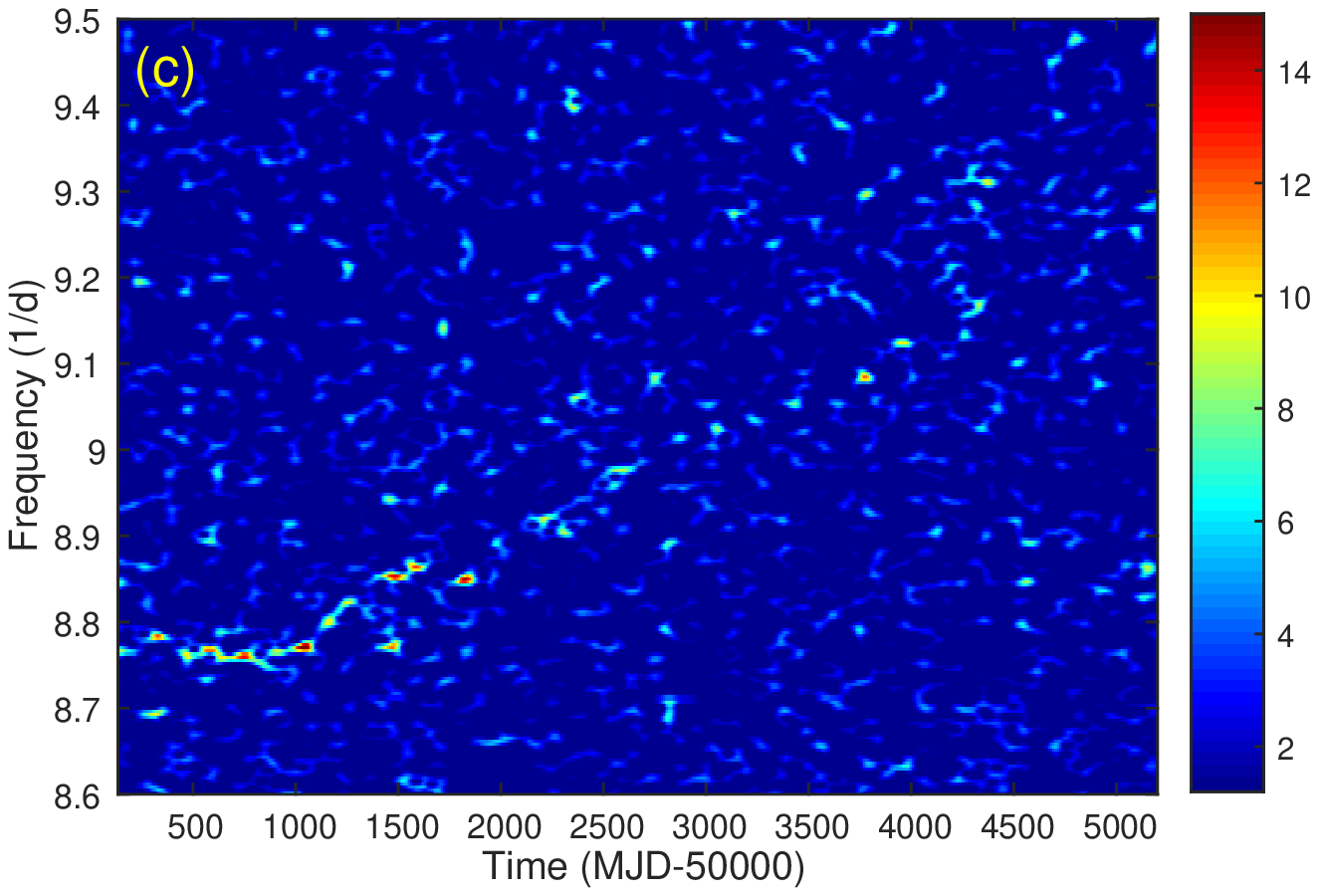}
\end{minipage}
\hspace{0.5cm}
\begin{minipage}{0.47\linewidth}
\includegraphics[width=0.95\textwidth]{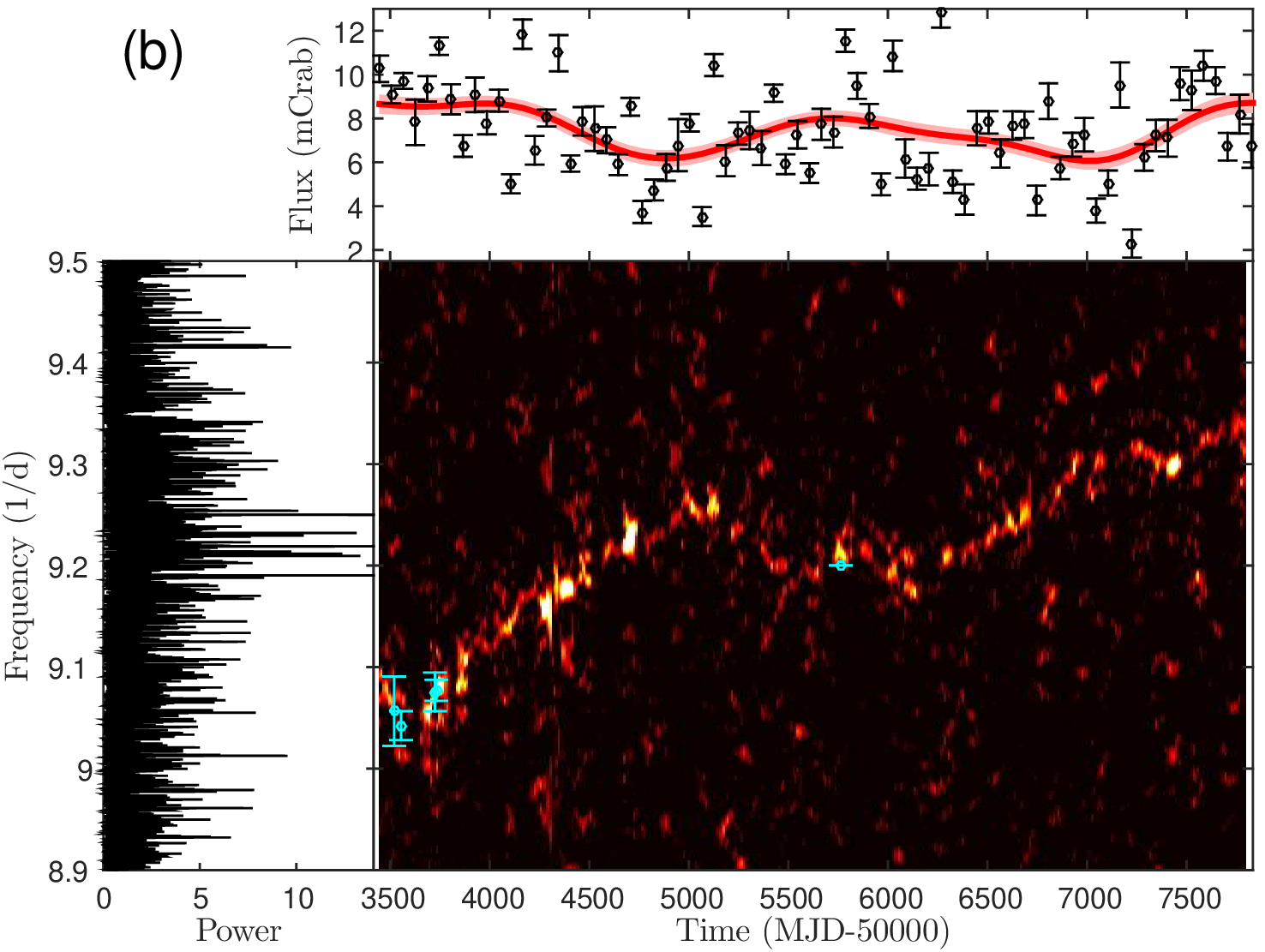}
\hspace{0.5cm}
\includegraphics[width=0.95\textwidth]{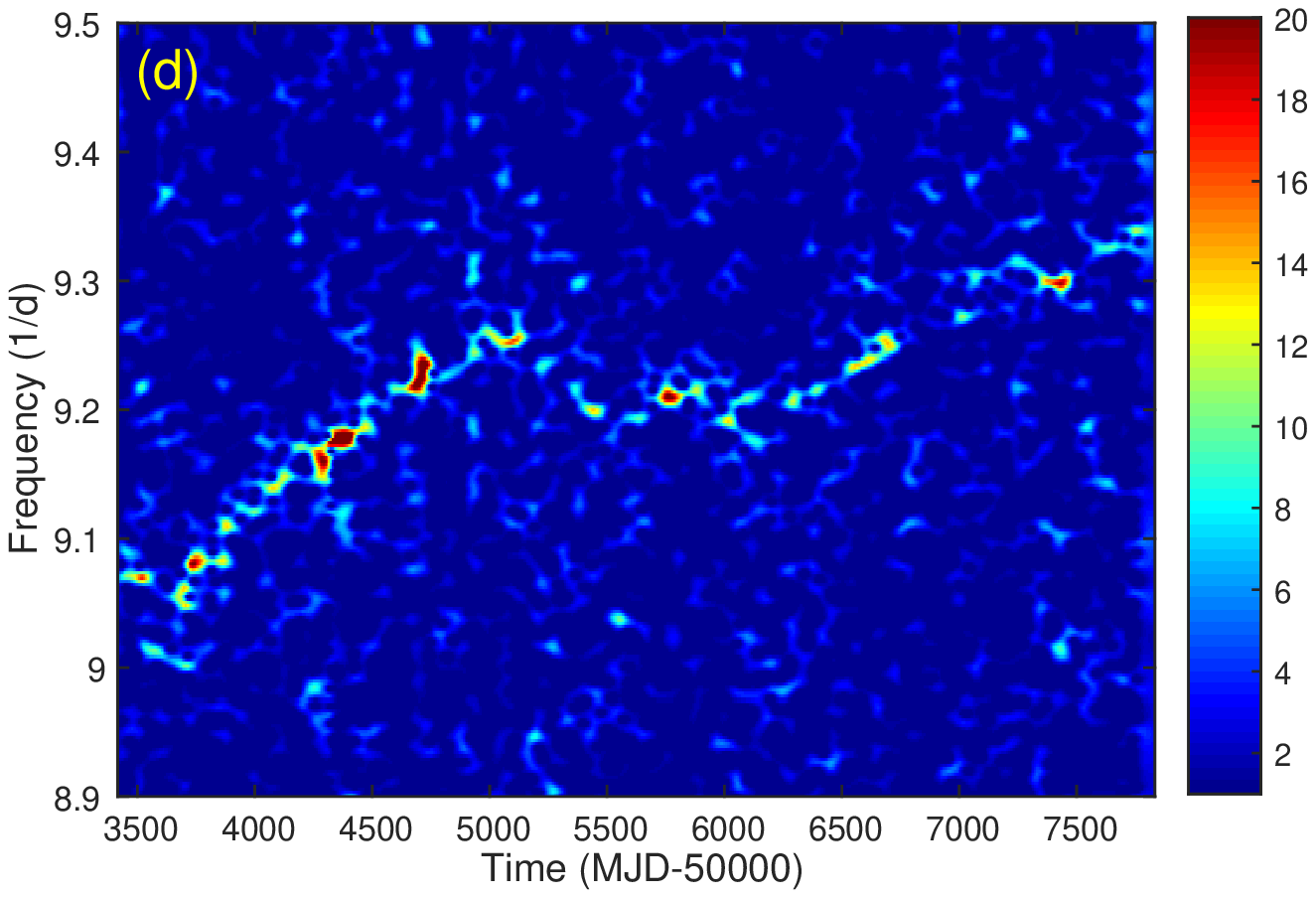}
\end{minipage}
\caption{\BD{(a)--(b) DPSs of spin frequency from (a) \rxte\ ASM and (b) \swift\ BAT light curves. Cyan points are previously reported spin frequencies from the observations obtained using \rxte\ PCA, \integral, and \suzaku. The left panels show the Lomb--Scargle power spectra of the entire data. The upper panels show the rebinned light curve (black data points), and the Gaussian smoothed low-passed filtered light curve (red lines).  The 90\% confidence levels for the Gaussian low-pass filtered light curve are labeled by the pink-shaded area. (c)--(d) WWZ spectra from (c) \rxte\ ASM and (d) \swift\ BAT light curves.} \label{dps_wwz_spin}}
\end{figure*}



To study the variability of the spin period, we created the dynamic power spectra (DPSs) using the Lomb--Scargle technique \citep{Scargle1982, Press1989} from the ASM and BAT data \citep[see, e.g., ][]{ClarksonCC2003a}.  We fixed the size of the moving window to 60 days, which is approximately two superorbital cycles, in order to maintain sufficient amount of data points and ensure that the spin period does not change significantly during the interval. The moving step was fixed to 10 days to check whether the spin period changed gradually.  We also generated a light curve rebinned into 60 days to observe the relation between the evolution of the spin period and the X-ray flux of \xb. The count rates are normalized to the unit of the Crab Nebula using the X-ray photons of Crab collected in the same time bin. Although the Crab flux remained stable, the flux of \xb\ jumped significantly even between the neighboring bins, which might be due to the intrinsic short-term variability. To obtain a long-term evolutionary trend of the X-ray flux, we used a low-pass filter based on the Fourier transform to filter these modulations with a time scale shorter than approximately three years. We further obtained a 90\% confidence interval of the filtered light curves by performing $10^4$ Monte Carlo simulations. We also used another independent method, the ensemble empirical mode decomposition \citep{Huang1998, Wu2009}, to perform the low-pass filter, and the result is consistent with the result from the Fourier transform.

\BD{We also employed another time-frequency analysis technique, the weighted wavelet z-transform (WWZ), to explore the time-frequency properties of \xb\ \citep{Foster1996}. The WWZ is based on the Morlet wavelet algorithm, in which the wavelet is composed of a sinusoidal wave windowed by a Gaussian envelope \citep{GrossmannM1984}.} \BD{Compared to the Morlet wavelet, the WWZ is optimized for investigating the unevenly sampled data. Similar to the DPS, the size of the wavelet is controlled by adjusting the width of the Gaussian envelope. Moreover, the WWZ adaptively adjusts the cycle length of the sinusoidal function in the wavelet according to the trial frequency. Taking the DPS as an analogy, the window size in the WWZ can be regarded as a well-defined number of waves with an adjustable cycle length instead of a fixed time interval.  We tuned the size of the wavelet to approximately the same as the window size in the DPS for the spin frequency range in order to perform a fair comparison. The WWZ is defined as follows:
\begin{equation}
Z=\frac{(N_{\textrm{eff}}-3)V_y}{2(V_x-V_y)}\rm{,}
\end{equation}
where $N_{\textrm{eff}}$ is the effective number of data points in a specific time, $V_x$ is the weighted variation of the data, and $V_y$ is the weighted variation of the model functions that adjust the wavelets to the trial frequency \citep{Foster1996}}. 

The resulting DPS of the \rxte\ ASM light curve is shown in Figure \ref{dps_wwz_spin}a. We also show the previously detected spin frequencies \BD{(listed in Table \ref{spin_point_obs})} in this figure.  The Lomb--Scargle power spectrum of the entire data is shown for comparison. The power spectrum is noisy, and no significant signals can be detected, indicating that the spin period is not steady.  On the other hand, an evolutionary track of the spin frequency appears on top of the noisy background in the DPS.  \BD{The WWZ spectrum is shown in Figure \ref{dps_wwz_spin}c, in which the evolutionary track is fully consistent with that in the DPS.} Before MJD 51000, we observe a marginal spin-down trend with a period derivative $\dot{P}_{\rm{spin}}=(4\pm3)\times 10^{-7}$ s s$^{-1}$ estimated by fitting the peaks of the DPS with a linear function.  When estimating the period derivative, we fixed the size of the moving step equal to the size of the moving window to make the frequency measurements independent of each other. This track is consistent with the previously reported detection using the ASM data \citep{CorbetFP1999}. During the same epoch, the flux increased with time.  Then, the DPS showed a long spin-up track, although the signal cannot be recognized after $\sim$ MJD 52500, possibly owing to the insufficient sensitivity of the ASM. Between MJD 51000 and 52500, the spin-up rate was $\dot{P}_{\rm{spin}}=(-1.50 \pm 0.07) \times 10^{-6}$ s s$^{-1}$. The X-ray flux during this epoch gradually decreased. Another remarkable feature is that it seems as if there exist a short-term random-walk/spin-down epoch between MJD 51500 and 52000, although the fluctuation in the long-term X-ray flux cannot be observed.

The DPS of the BAT light curve was generated in the same manner as that of the ASM, as shown in Figure \ref{dps_wwz_spin}b.  The Lomb--Scargle power spectrum for the entire data set does not exhibit a significant peak dominating the spectrum; however, a concentration of weak signals with $f\approx9.2-9.3$\,d$^{-1}$ can be observed. Compared with the ASM DPS, the evolution of the spin frequency is more clearly detected \BD{owing to a better sensitivity}. The evolution of the spin period was complex and the significance of the signal also dramatically varied with time. \BD{The WWZ spectrum presented in Figure \ref{dps_wwz_spin}d also shows a consistent evolutionary pattern compared with the DPS.} Two long spin-up trends were clearly observed in MJD 53700--55000 and 56000--57000 with $\dot{P}_{\rm{spin}}=(-1.68 \pm 0.08) \times 10^{-6}$ s\,s$^{-1}$ and $\dot{P}_{\rm{spin}}=(-1.7 \pm 0.1) \times 10^{-6}$ s\,s$^{-1}$, respectively. In addition, another long spin-down/random-walk epoch with $\dot{P}_{\rm{spin}}=(6 \pm 2) \times 10^{-7}$ s\,s$^{-1}$ occurred in MJD 55000--56000. \BD{After $\sim$MJD 57000, the secular spin-up trend seems to be broken, but we cannot estimate its behavior until a new spin-up trend is established.}  The spin periodicities detected with the \rxte\ PCA, \integral, and \suzaku\ (listed in Table \ref{spin_point_obs}) are in good agreement with the BAT result. 

All the spin period derivatives of these five spin evolution epochs and the corresponding flux trends are listed in Table \ref{spin_epoch}. If we fit the evolution of the spin period with the entire \emph{Swift} data set by a straight line, a period derivative $\dot{P}_{\rm{spin}}=(-5.6\pm 0.6)\times 10^{-7}$ s s$^{-1}$ can be obtained.  In general, the pulsation signal appears intermittently and shows significant fluctuation.  In addition to the long-term trend, the spin period evolution also showed fluctuations on much shorter time scales. Torque switches are clearly observed in the spectra, and the spin frequency seems to be anticorrelated with the X-ray flux.

\begin{deluxetable}{ccc}
\tablecaption{Five spin evolution epochs of \xb. \label{spin_epoch}} 

\tablehead{\colhead{Time Interval} & \colhead{$\dot{P}_{\rm{spin}}$} & Flux Trend \\ 
\colhead{(MJD-50000)} & \colhead{($10^{-7}$s\,s$^{-1}$)} &} 
\startdata
0--1000 & $4\pm3$ & Increase \\ 
1000--2500 & $-15.0\pm0.7$ & Decrease\\
3700--5000 & $-16.8\pm0.8$ & Decrease\\
5000--6000 & $6\pm2$ & Increase\\
6000--7000 & $-17\pm1$ & Decrease\\
\tableline
3500 -- 7700 & $-5.6\pm0.6$ & \nodata 
\enddata
\end{deluxetable}

\begin{figure*}[t]
\begin{minipage}{0.47\linewidth}
\includegraphics[width=0.95\textwidth]{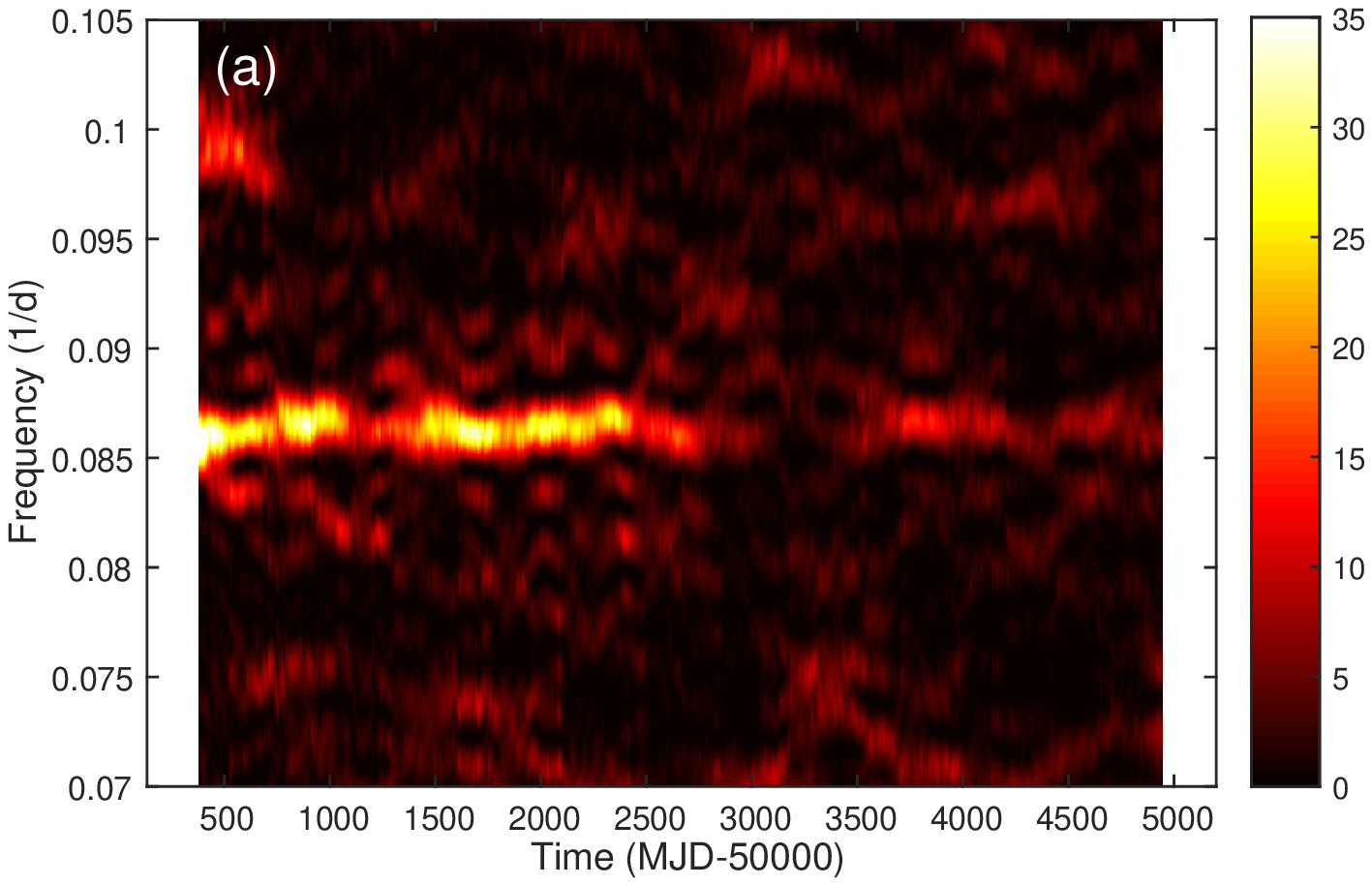}
\includegraphics[width=0.95\textwidth]{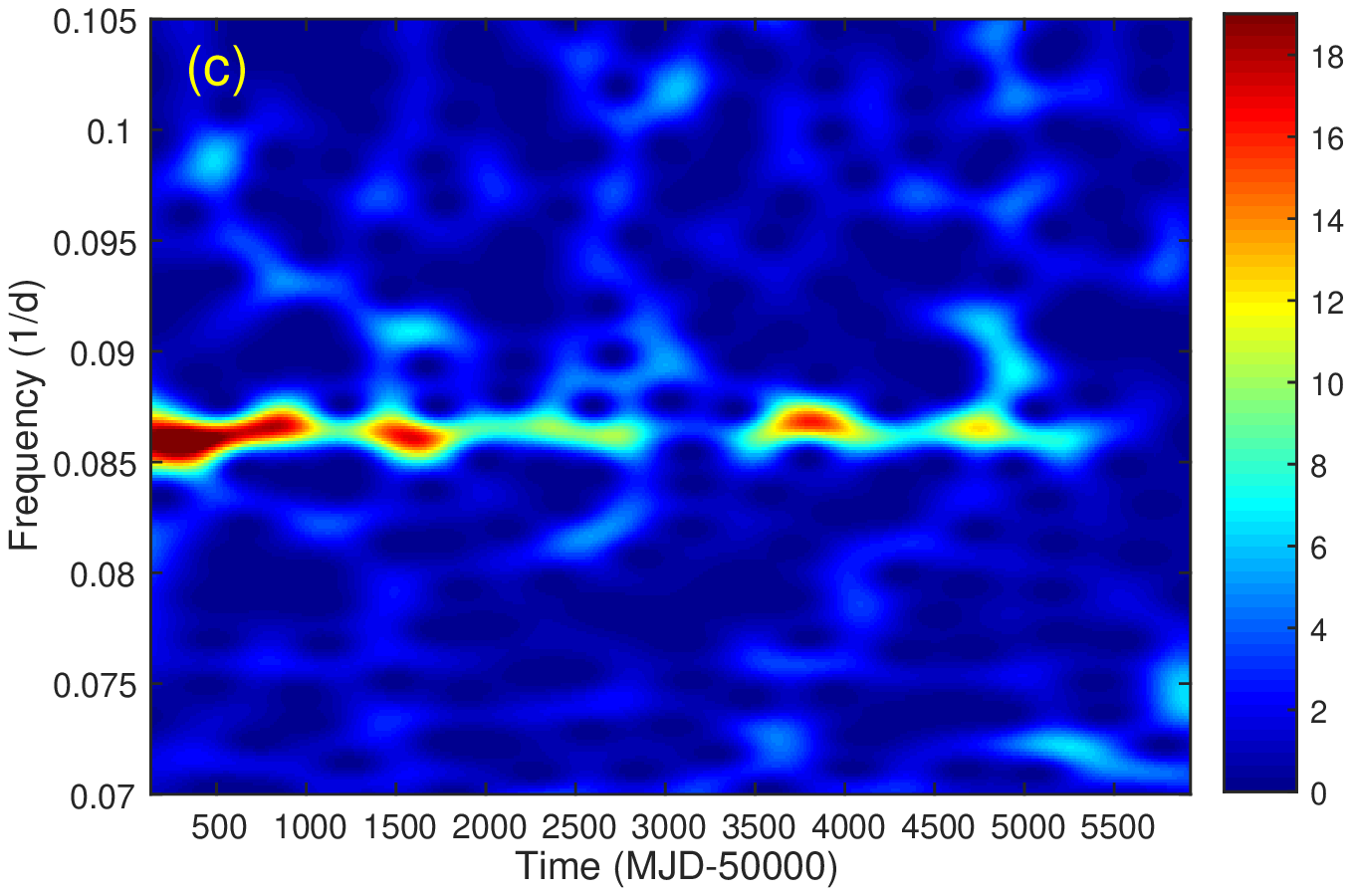}
\end{minipage}
\hspace{0.5cm}
\begin{minipage}{0.47\linewidth}
\includegraphics[width=0.95\textwidth]{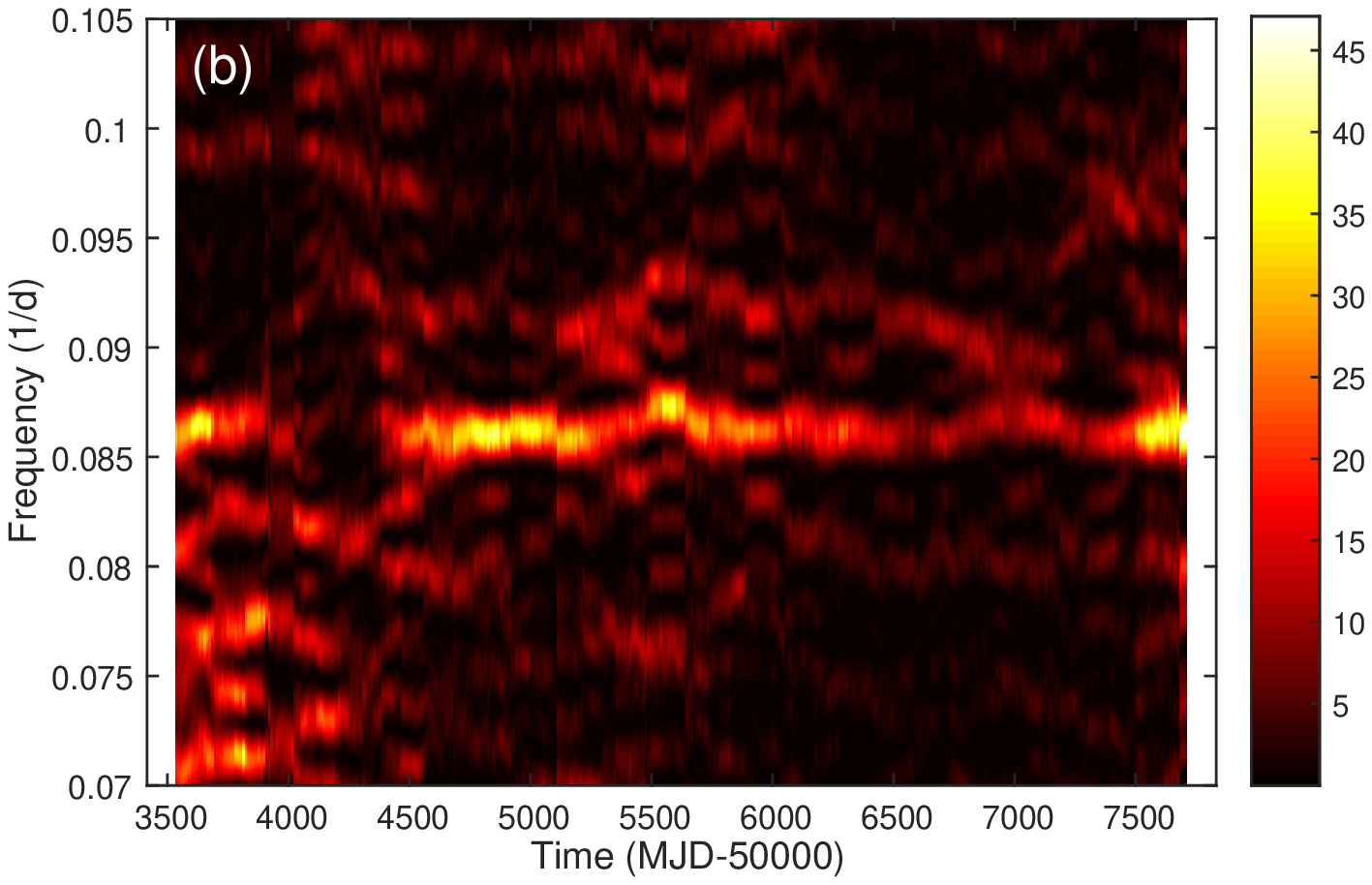}
\includegraphics[width=0.95\textwidth]{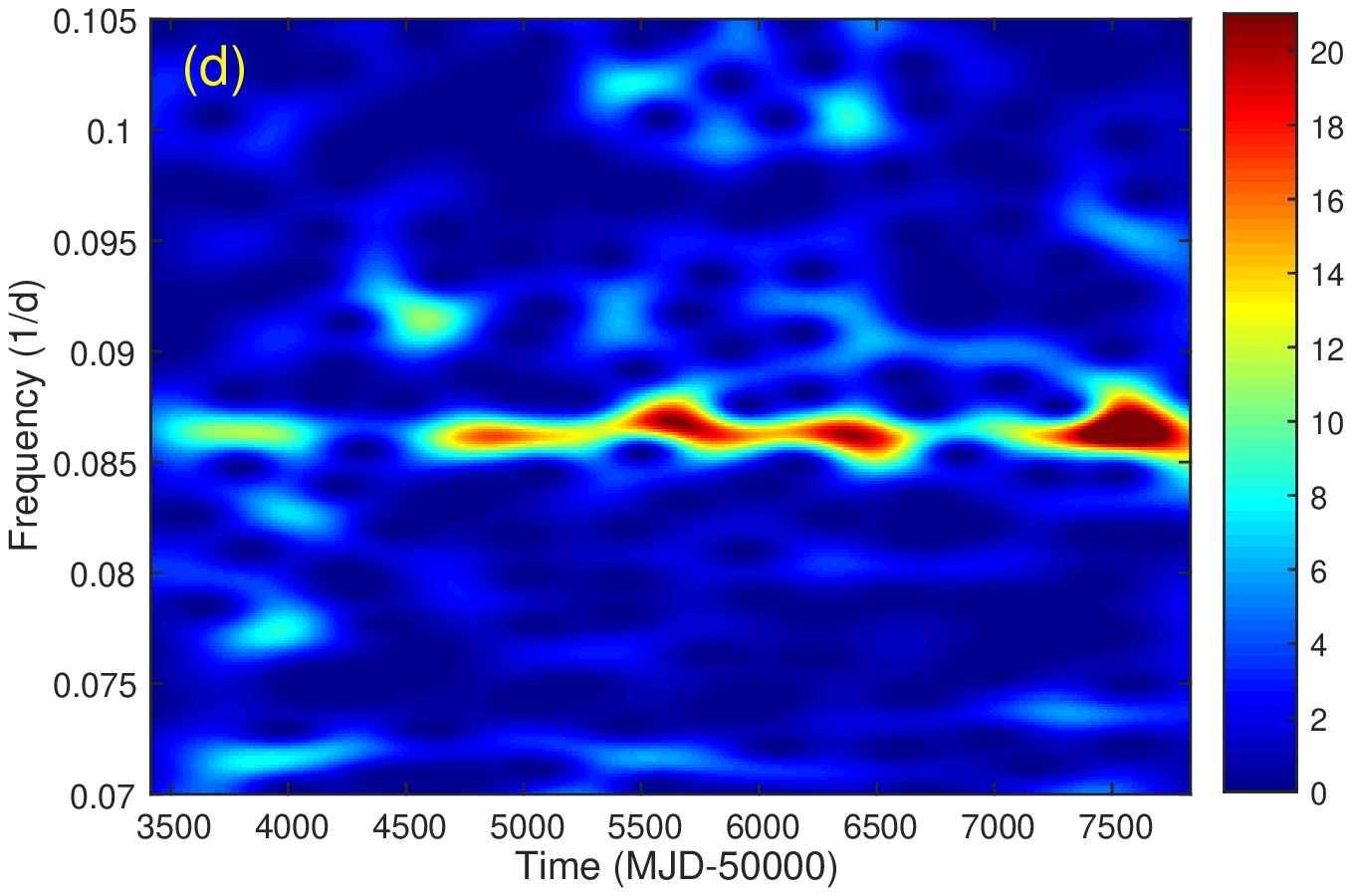}
\end{minipage}
\caption{\BD{(a)--(b) DPS of (a) ASM and (b) BAT light curves in the frequency range of the orbital period. (c)--(d) WWZ spectra of (c) ASM and (d) BAT light curves in the same frequency range.} \label{dps_orbital}}
\end{figure*}


\begin{figure}
\centering
\plotone{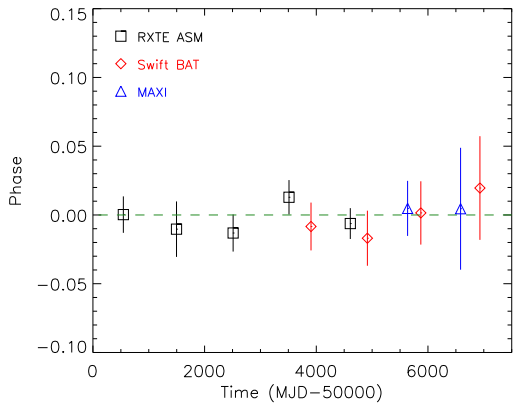}
\caption{Phase evolution of orbital modulation. The black squares, red diamonds, and blue triangles are dip arrival phases obtained with \rxte\ ASM, \swift\ BAT, and MAXI, respectively.   \label{o_c_orbital}}
\end{figure}

\begin{figure*}
\centering
\plotone{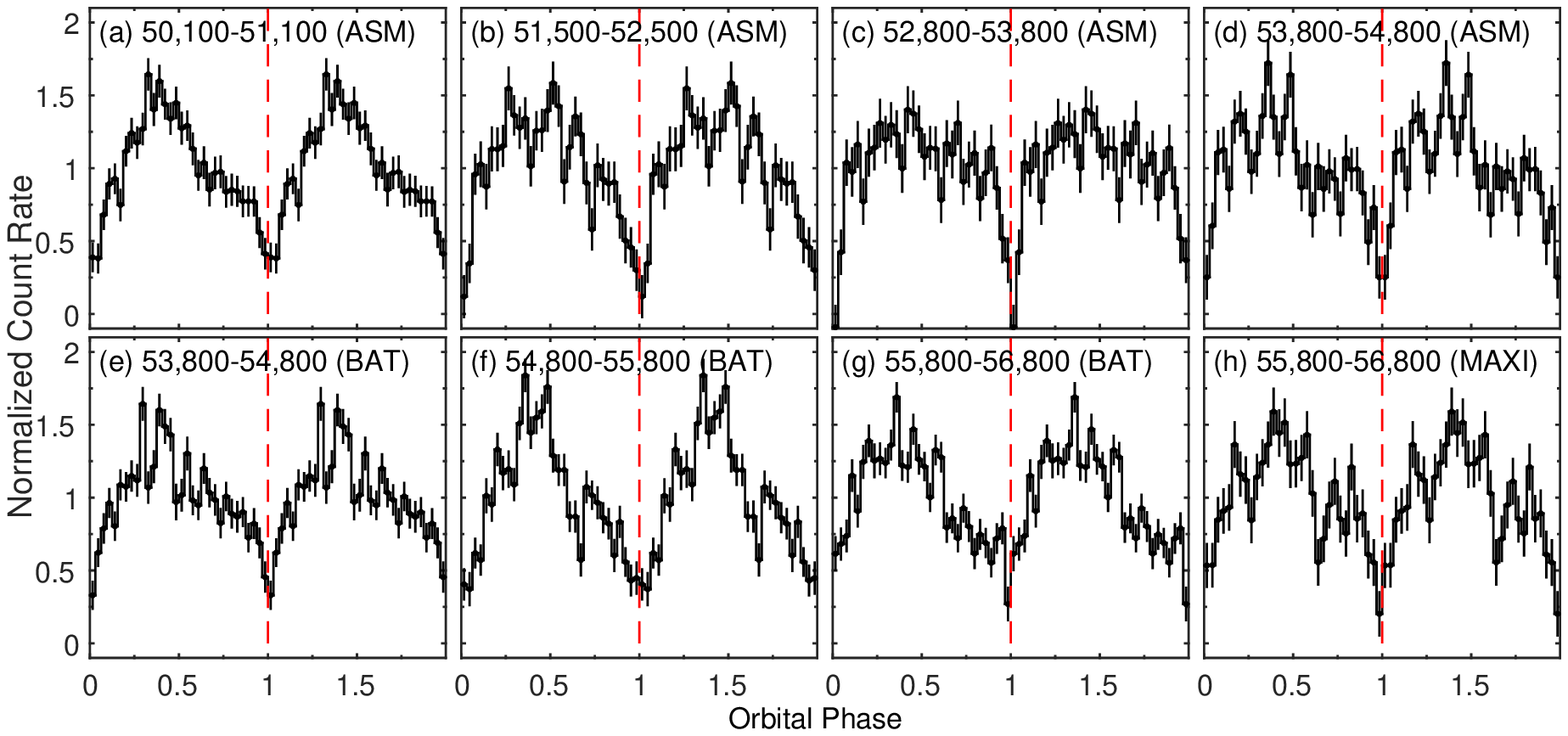}
\caption{Typical orbital profiles obtained in different epochs with (a)--(d) \rxte\ ASM, (e)--(g) \swift\ BAT, and (h) MAXI . The $y$-axes are the normalized count rates and the $x$-axes are the orbital phases.   \label{fold_lc_orbital}}
\end{figure*}

\subsection{Orbital Modulation}\label{evo_orbital}
The orbital modulation can be detected using the all-sky monitoring programs \citep{CorbetFP1999, WenLC2006, PradhanPP2015}. We used the DPS and WWZ to further examine the stability of the orbital period with a longer time span.  Owing to the faint nature of \xb, we fixed a window size of 500 days with a moving step of 10 days. \BD{A shorter window size can reveal a more detailed variability of the signal, but the frequency resolution will be poorer and the contribution of the noise is more significant. For this reason, we only discuss variabilities longer than $\sim500$~days in this paper.} The DPSs and WWZs for the ASM and BAT datasets are shown in Figure \ref{dps_orbital}.  There is no clear variation in the orbital period; however, both the Lomb--Scargle power and the WWZ strength changed remarkably with time.  The signal was significantly detected in the first $\sim$1,000 days of the ASM observations, and the power obviously \BD{dropped} because of the decrease in the X-ray flux of \xb. In addition to this long-term trend, some short-term variabilities of the power are observed, e.g., the signal became weak during MJD 52800 -- 53500.  To investigate the variation in the orbital profile and the detailed phase evolution, we divided the light curves into several segments and fit the dip feature with a Gaussian function to determine the flux minimum as the fiducial point for the phase analysis. We included the MAXI light curve in this analysis to check the consistency between the hard X-ray profile determined using the \swift\ BAT and the soft X-ray profile determined with the MAXI. The best-fit linear ephemeris for describing the intensity minima of the dip feature is provided as follows:
\begin{equation}
T_N=\rm{MJD } (50126.99\pm0.09) + (11.5979\pm0.0003)\times N\rm{,}
\end{equation}
where $T_N$ is the $N$th expected dip arrival time (see Figure \ref{o_c_orbital}).  This period is fully consistent with the period determined with three years of optical spectroscopy \citep{GrundstromBG2007}. The X-ray minima in our ephemeris correspond to the orbital phase $0.62\pm0.05$ of the optical radial velocity, which is in good agreement with the predicted supergiant inferior conjunction at phase $0.65\pm0.06$ \citep{GrundstromBG2007}.  

In addition, we found that the dip was not always clearly shown, and the sawtooth modulation profile was also highly variable.  Figure \ref{fold_lc_orbital} shows some examples of the folded light curve with a time interval of 1000 days. The orbital profile in Figure \ref{fold_lc_orbital}(a) is extracted from the first $\sim$1000 days of the \rxte/ASM light curve, when the source was bright.  Both the sawtooth modulation and the dip are obvious.  However, during this epoch, the dip was the least visible one compared with other epochs in the ASM observations.  From Figure \ref{fold_lc_orbital}(a) to (c), we found that the sawtooth feature decreased with time and was almost invisible between MJD 52800 and MJD 53800 (Figure \ref{fold_lc_orbital}(c)). The sawtooth modulation was almost invisible, and only the dip can be observed. The decrease in power in this time epoch was caused by the insensitivity of both the Lomb--Scargle and the WWZ algorithms for the highly non-sinusoidal modulation shape. After MJD 53800, the sawtooth modulation reappeared (see Figure \ref{fold_lc_orbital}(d)). 

Figure \ref{fold_lc_orbital}(e) shows the orbital profile detected with \swift\ BAT in the same time epoch as Figure \ref{fold_lc_orbital}(d).  In addition to the sawtooth modulation caused by the stellar wind, the dip was also clearly observed.  Furthermore, the dip obtained with \swift\ BAT was less obvious than the dip detected with \rxte\ ASM, indicating that this feature is likely caused by absorption.  Interestingly, the dip is blurred in Figure \ref{fold_lc_orbital}(f) but the flux minimum is still located at phase 0. At the same time, the sawtooth profile is obvious. This epoch, together with the epoch in Figure \ref{fold_lc_orbital}(a), coincides with the two long spin-down/random-walk epochs described in Section \ref{evo_spin}. Moreover, the Lomb--Scargle power of the orbital modulation during these two epochs is higher than other epochs. Then, the dip seemed to reemerge in the final data segment of the \swift\ observations (Figure \ref{fold_lc_orbital}(g)) although the width was extremely narrow. \BD{The orbital profile obtained with MAXI in the same time epoch between MJD 55800 and MJD 56800 is shown in Figure \ref{fold_lc_orbital}(h), which confirms the detection of the dip.}

\subsection{Superorbital Modulation}
\begin{figure*}[ht]
\begin{minipage}{0.47\linewidth}
\includegraphics[width=0.95\textwidth]{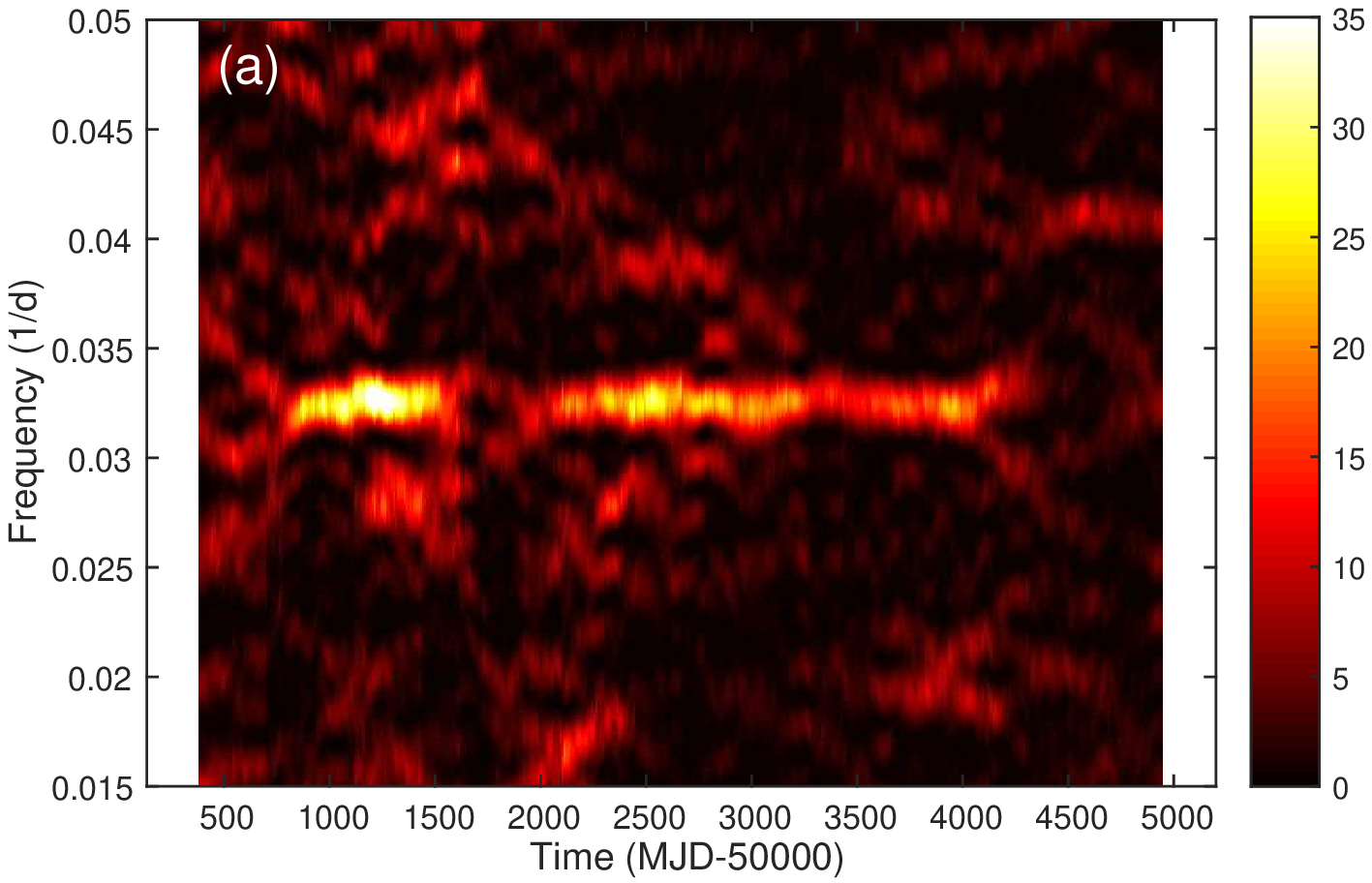}
\includegraphics[width=0.95\textwidth]{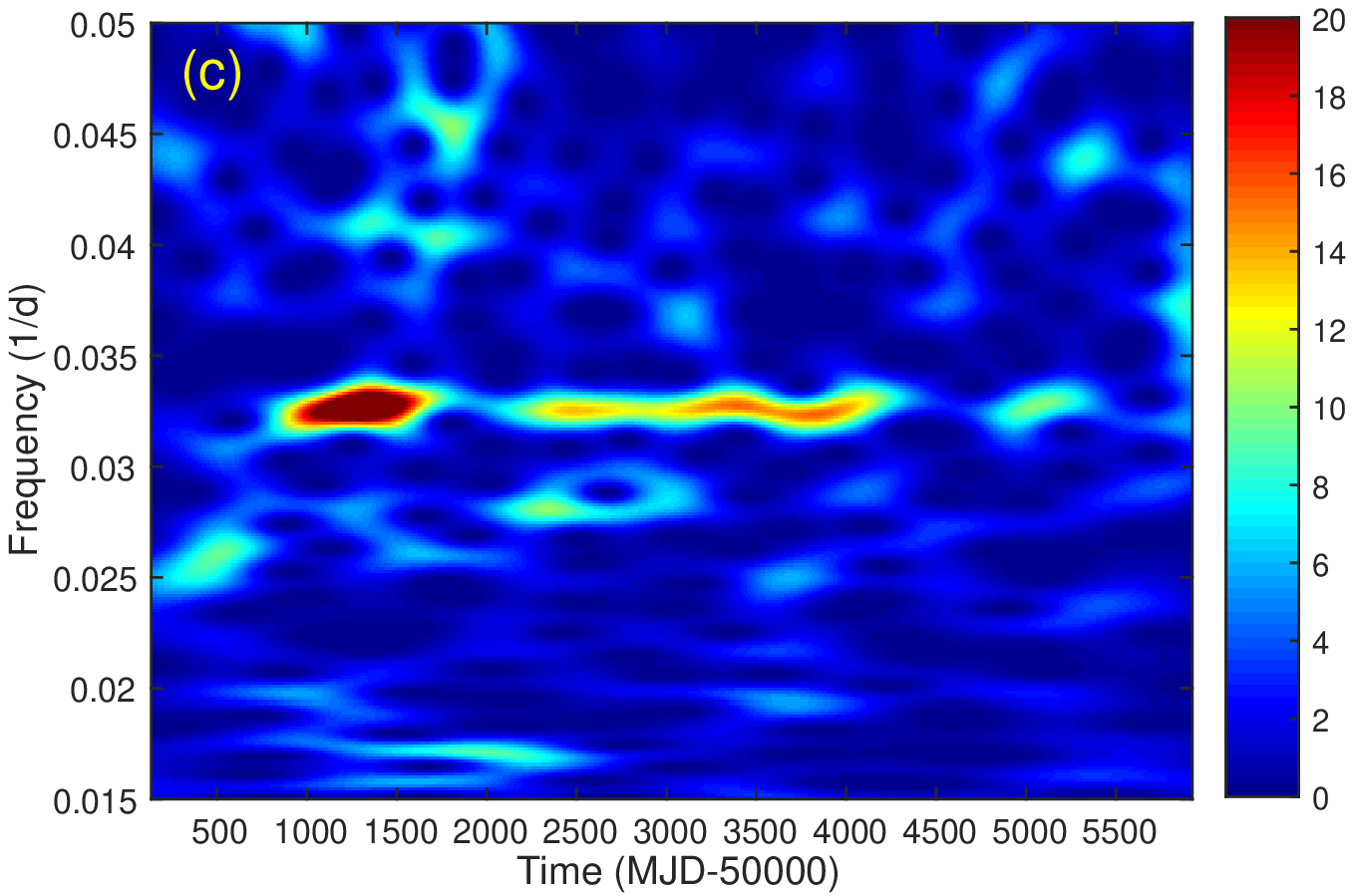}
\end{minipage}
\hspace{0.5cm}
\begin{minipage}{0.47\linewidth}
\includegraphics[width=0.95\textwidth]{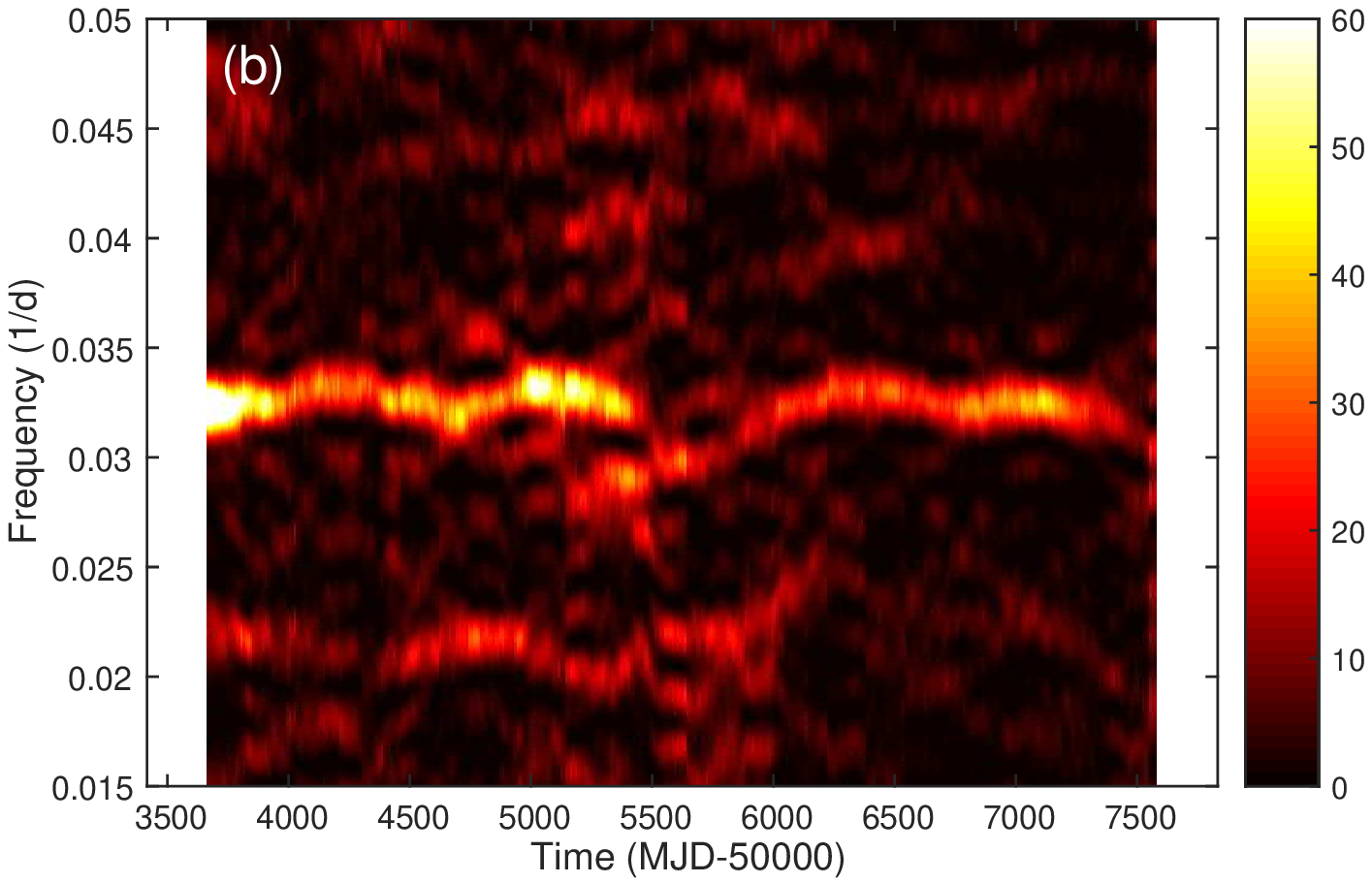}
\includegraphics[width=0.95\textwidth]{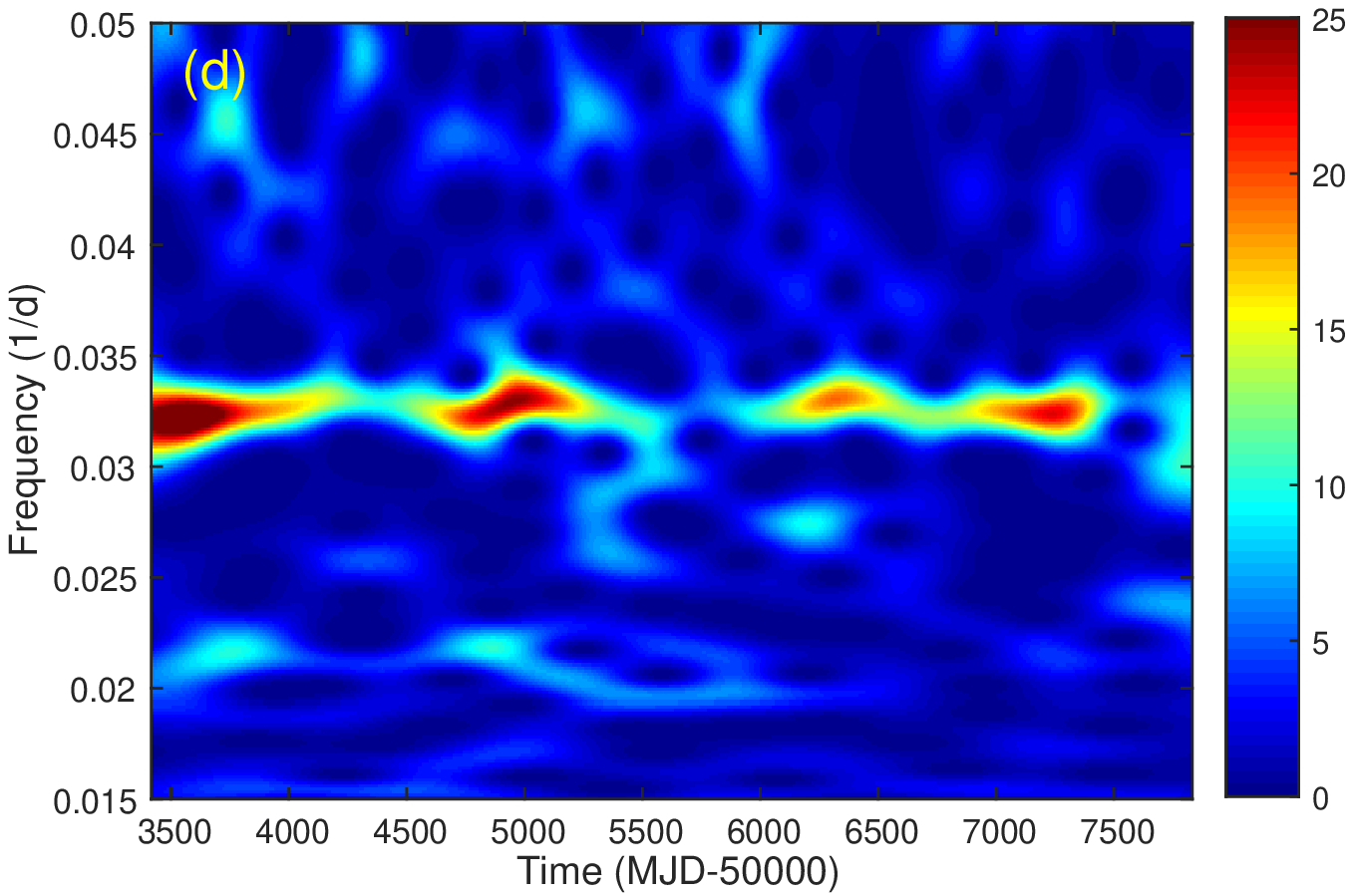}
\end{minipage}
\caption{\BD{(a)--(b) DPS of (a) ASM and (b) BAT light curves in the frequency range of the superorbital period. (c)--(d) WWZ spectra of (c) ASM and (d) BAT light curves in the same frequency range.} \label{dps_superorbital}}
\end{figure*}

The physical origin of the superorbital modulation of \xb\ is still unclear, and previous works have confirmed that the period was stable \citep{FarrellSO2006, KotzeC2012}. Because \rxte\ ASM and \swift\ BAT light curves extended to a much longer time baseline, we investigated the evolution of the superorbital modulation with both the DPS and the WWZ techniques. We produced DPSs with a window size of 500\,days and a moving step of 10\,days. The results are shown in Figure \ref{dps_superorbital}a and \ref{dps_superorbital}b, in which the period is still stable, but the power is highly variable with time. \BD{The WWZ spectra are shown in Figures \ref{dps_superorbital}c and \ref{dps_superorbital}d, and the characteristics are consistent with the DPSs.} There are several weakly detected epochs in the ASM dataset: before $\sim$ MJD 51000, $\sim$ MJD 51500--52000, and after $\sim$ MJD 54000.  The last one may be caused by the insufficient sensitivity of the ASM because the signal was significantly detected in the \swift\ BAT dataset.  The DPS of the BAT light curve also shows a break of the superorbital modulation signal in $\sim$ MJD 55300 -- 56000.  These decreases in the superorbital signal apparently coincide with the presence of spin-down/random-walk epochs, although the length and time are slightly offset. 

\begin{figure*}[t]
\centering
\begin{minipage}{0.47\linewidth}
\includegraphics[width=0.95\columnwidth]{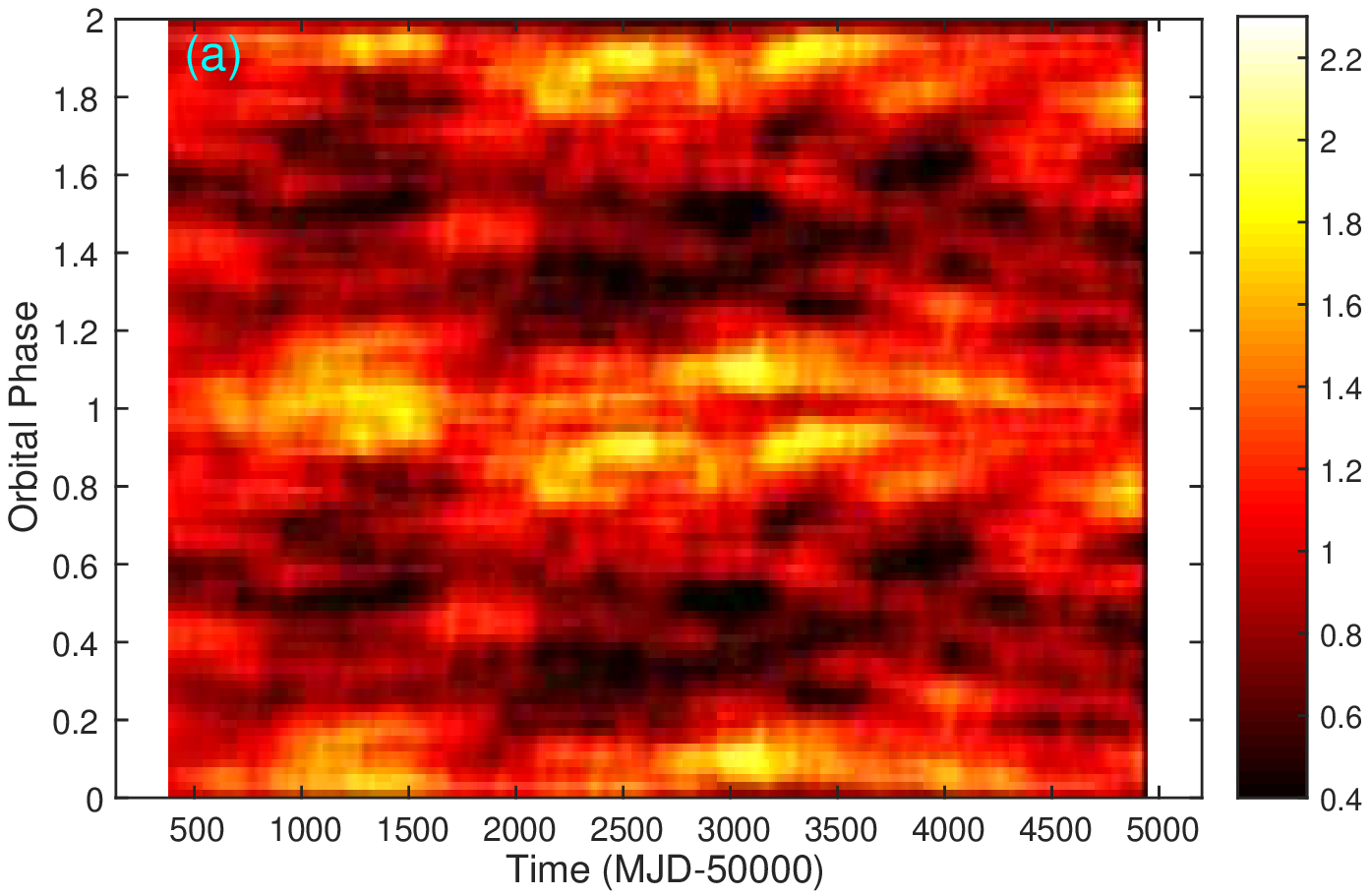}
\end{minipage}
\hspace{0.5cm}
\begin{minipage}{0.47\linewidth}
\includegraphics[width=0.95\columnwidth]{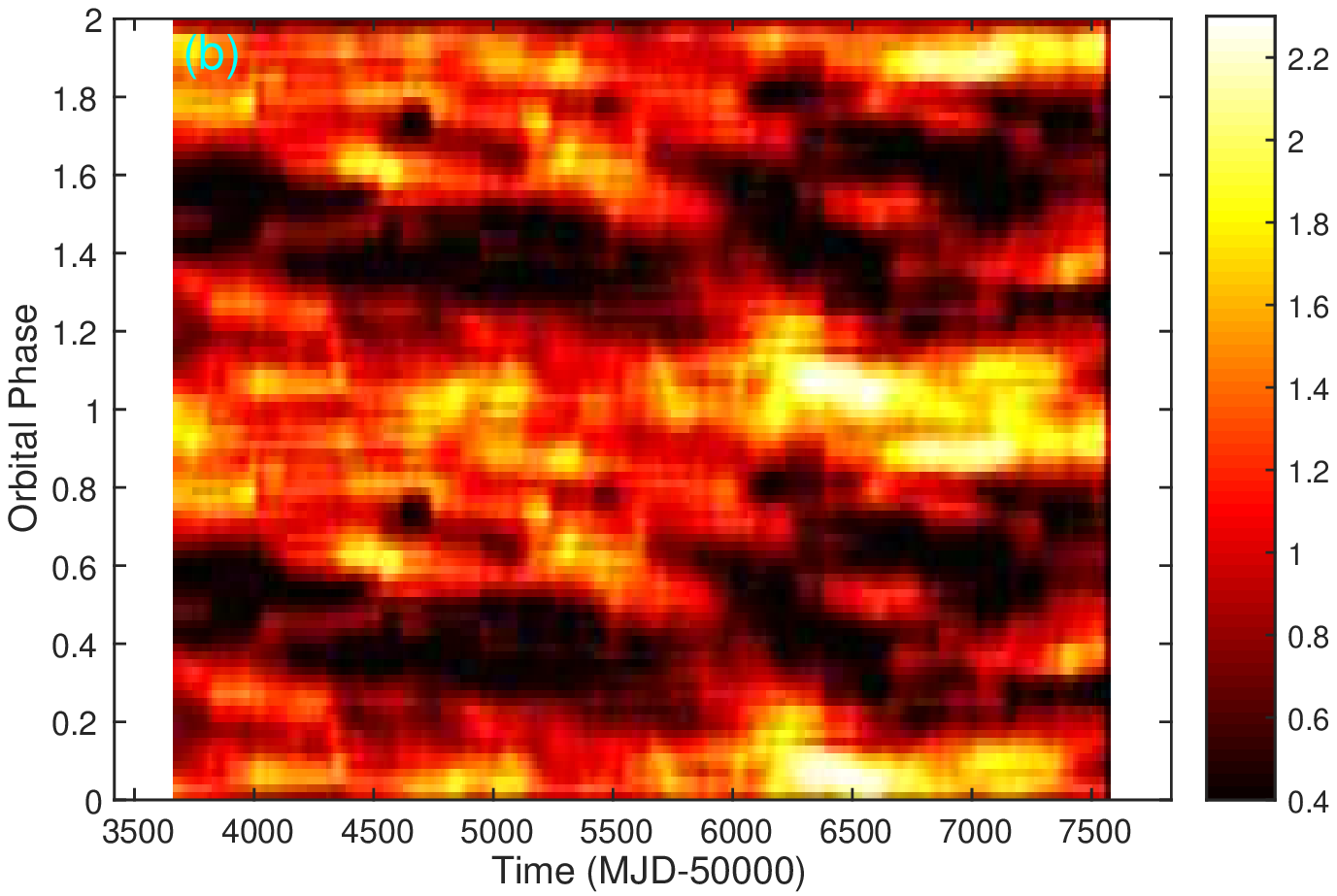}
\end{minipage}
\caption{Dynamic superorbital profiles obtained from ASM (a) and BAT (b) data set. The colors represent the normalized count rate.  \label{dfl_superorbital}}
\end{figure*}

To explore the connection between the superorbital modulation signal and the spin behavior, we further examined the variability of the superorbital modulation profile.  Analogous to the DPS, we created the ``dynamically folded light curve.''  Instead of calculating the Lomb--Scargle power spectrum, we generated a normalized folded light curve according to the superorbital modulation period in each moving window, collected all the folded light curves, and present the variability of superorbital modulation profile in a color map (see Figure \ref{dfl_superorbital}).  The superorbital modulation is not clearly recognized in the first two years of the ASM data, as indicated by \citet{FarrellSO2006}. Then, the superorbital modulation profile showed a sharp peak and a flat valley.  During $\sim$MJD 51500 and MJD 52000, the superorbital modulation profile was blurred again, which corresponds to the weak detection in the DPS (see Figure \ref{dps_superorbital}a). Then, the width of the peak increased with time until the end of the ASM data.  Owing to better sensitivity, the DPS of BAT during the overlapping epoch was much more significant than that of the ASM although the modulation shape was consistent.  During $\sim$ MJD 55500 and $\sim$ MJD 56200, the modulation profile was blurred similar to that in the first two years of the ASM data.  After MJD 56200, the sharp peak reappeared like the behavior in $\sim$MJD 51000.  If the variation in the superorbital modulation profile is recurrent, the time scale would be approximately 5000 days.

\begin{figure}
\centering
\plotone{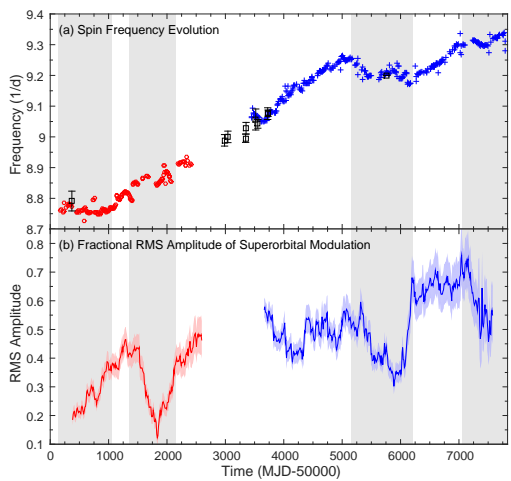}
\caption{(a) Evolution of the spin frequency of \xb.  The red circles are obtained with \rxte\ ASM and the blue plus signs are determined with \swift\ BAT. The gray-shaded area denotes the random-walk epochs, while the spin frequency is increasing in other epochs.   (b) The variation in the RMS amplitude of superorbital modulation profile obtained with \rxte\ ASM (red) and \swift\ BAT (blue).  \label{spinfreq_rmsamp}}
\end{figure}

We further calculated the fractional RMS amplitude of the superorbital modulation following the method of \citet{VaughanEW2003}. With the same analysis as the DPS and the dynamically folded light curve, we obtained the fractional RMS amplitudes within a 500-day data window, and the variation is shown in Figure \ref{spinfreq_rmsamp}.  The spin frequency evolution is also plotted in the same figure for comparison. We show the spin signal from \rxte\ ASM only before $\sim$MJD 52500 because it cannot be significantly determined later. It is clear that the RMS amplitudes of the superorbital modulation dropped during the spin-down/random-walk epochs.

\section{Discussion}\label{discussion}
\subsection{Nature of the Spin Evolution}
Assuming that the accreting materials from a prograde disk transfer all the angular momentum to the accreting neutron star, the change in spin period is dominated by the accretion torque:
\begin{equation}
N\approx \dot{M} \sqrt{GM_{\rm{NS}}r_m},
\end{equation}
where $\dot{M}$ is the mass accretion rate, $M_{\rm{NS}}$ is the mass of the neutron star, and $r_m$ is the magnetospheric radius \citep{RappaportJ1977,BildstenCC1997}. When the magnetospheric radius is smaller than the corotating radius $r_{\rm{co}}=(GM_{\rm{NS}}P_{\rm{spin}}^2/4\pi^2)^{1/3}$, the pulsar gains a positive angular momentum and has a spin-up rate as follows:
\begin{equation}
\begin{split}
\dot{P}_{\rm{spin}}=& -1.6\times10^{-13} \left( \frac{\dot{M}}{10^{-10}M_{\odot}\rm{yr}^{-1}} \right)\\
 & \times \left( \frac{P_{\rm{spin}}}{s} \right)^{7/3} \left( \frac{r_{\rm{co}}}{r_m} \right)^{1/2}\textrm{,}
\end{split}
\end{equation}
where $P_{\rm{spin}}$ is the spin period. Otherwise, the accreting materials will be ejected out through the magnetic field lines and a spin-down torque applies on the pulsar via the propeller effect when $r_{\rm{co}}<r_m$ \citep{IllarionovS1975}. On the other hand, the accretion torque in a wind-fed system is random and the spin period evolution shows a random-walk behavior, which is the accumulation of white noise and has no long-term trend in the spin period evolution \citep{Corbet1986}.  Vela X-1 is a typical purely wind-fed system that shows random walk in pulsation frequency \citep{DeeterBL1989}.

However, this simple picture was changed with the long-term short-cadence monitoring of numerous pulsars with BATSE \citep{BildstenCC1997} and \fermi\ GBM \citep{FingerBN2009}. For example, the disk-fed accreting pulsar Cen X-3 displayed rapid transitions between spin-up and spin-down with a much larger amplitude than the long-term torque. The spin history of the wind-fed accreting pulsar OAO 1657$-$415 was similar to that of Cen X-3, suggesting that it had an accretion disk \citep{FingerBN2009}. A more detailed study of the flux-torque correlation suggested that OAO 1657$-$415 switches between the direct wind accretion and the disk-wind accretion \citep{JenkeFW2012}. During the direct wind accretion epoch, no disk formed around the pulsar and the torque from the accreting material is uniform and stochastic. In the disk-wind accretion phase, a disk formed and a strong torque is applied on the neutron star and causes a secular spin-up trend. However, the mechanism triggering the disk formation is still unknown. Another wind-fed system GX 301$-$2 was considered to exhibit a transient accretion disk, because it occasionally showed short-term dramatic spin-up epochs \citep{KohBC1997}. Moreover, the stochastic noise was observed in all accreting systems. These discoveries blurred the boundary between wind-fed and disk-fed systems.

Previous studies suggested that the spin-up rate of \xb\ was $\dot{P}_{\rm{spin}}/P_{\rm{spin}}\sim-2\times10^{-3}$\,yr$^{-1}$ \citep{HallFC2000, FarrellSO2008}, indicating a spin-up timescale of $\sim 400$\,years, which is longer than the timescale of $\sim 100$ years if \xb\ is driven by disk accretion \citep{Ikhsanov2007}. Thus, the wind-fed accretion scenario was favored \citep{FarrellSO2008}. The overall spin period derivative obtained from the BAT light curve is fully consistent with the spin-up rate in previous studies.  However, the spin-up epochs listed in Table \ref{spin_epoch} have significantly larger period derivatives.  The period derivative during the spin-up epochs is $\dot{P}_{\rm{spin}}/P_{\rm{spin}}\approx-6\times10^{-3}$\,yr$^{-1}$, yielding a much shorter spin-up timescale of $\sim150$\,years. At the same time, the X-ray flux showed decreasing trends. \BD{On the other hand, the period derivative during the spin-down/random-walk epochs is $\dot{P}_{\rm{spin}}/P_{\rm{spin}}\approx1.5\times10^{-3}$\,yr$^{-1}$. These two values are close to the mean values in the disk-wind accretion epochs ($\dot{P}_{\rm{spin}}/P_{\rm{spin}}\approx-7\times10^{-3}$\,yr$^{-1}$) and the direct wind accretion epochs ($\dot{P}_{\rm{spin}}/P_{\rm{spin}}\approx2\times10^{-3}$\,yr$^{-1}$) of OAO 1657$-$415, respectively \citep{JenkeFW2012}.} The long-term spin-up rate of \xb\ determined with \swift\ BAT ($\dot{P}_{\rm{spin}}/P_{\rm{spin}}\approx-1.8\times10^{-3}$\,yr$^{-1}$) is also comparable with that of OAO 1657$-$415 \citep[$\dot{P}_{\rm{spin}}/P_{\rm{spin}}\approx-1.1\times10^{-3}$\,yr$^{-1}$, ][]{ChakrabartyGP1993, JenkeFW2012}. These similarities indicate that \xb\ exhibits a transient or moderate-lived ($\sim1000-1500$ days) accretion disk the likely formed from the disk-wind accretion. The occasionally detected He{\sc ii} 4686\,\r{A} line suggested the presence of a transient accretion disk in \xb\ \citep{AabBK1983, CramptonHC1985, vanKerkwijkW1989}. Moreover, the hard X-ray tail, which is a signature linked to the presence of an accretion disk or a disk corona, is also occasionally detected in the \integral\ observations of \xb\ \citep{Wang2011}. Although the mechanism triggering the disk formation is unclear, the accretion disk can form when the circularization radius ($R_{\rm{circ}}$) is considerably larger than the size of the Neutron star. In the wind-fed system, 
\begin{equation}
R_{\rm{circ}}=\frac{2G^3M^3_{\rm{NS}}\omega^2_{\rm{orb}}}{v_{\rm{rel}}^8},
\end{equation}
where $\omega_{\rm{orb}}$ is the orbital angular frequency, and $v_{\rm{rel}}$ is the relative velocity between the neutron star and the stellar wind \citep{FrankKR2002}. Therefore, the formation of the accretion disk is likely triggered by the variation of the wind velocity.

According to the latest radial velocity measurement, the binary mass function of \xb\ is $f(m)=0.0032\pm0.0005$\,$M_\odot$ \citep{GrundstromBG2007}. Assuming a neutron star mass of 1.4\,$M_\odot$ and the best-fit inclination angle of 45$^\circ$, the mass of the supergiant can be calculated as $16\pm2$\,$M_\odot$. Furthermore, employing the approximation formula in \citep{Eggleton1983}, the Roche-lobe radius of the supergiant can be estimated as $33\pm2$\,$R_\odot$. Moreover, the radius of the supergiant was estimated as $37\pm15$\,$R_\odot$ from the optical photometry \citep{ReigCC1996}. This measurement is in the same order as the Roche-lobe size. Therefore, the transient Roche-lobe overflow is also possible if the radius of the companion varies with time. Both the disk-wind accretion and the Roche-lobe overflow are closely related to the activity of the companion supergiant.

Similar to the cases of Cen X-3 and OAO 1657$-$415, the spin evolution of \xb\ could contain a considerable number of short-term spin-up and spin-down epochs with random-walk noises.  The long-term spin-up trend may be contributed by more short-term spin-up epochs with the presence of an accretion disk.  In addition, the short-term spin-up epochs may also appear in the long-term spin-down/random-walk epochs, although the number should be much smaller. A long-term short-cadence monitoring of the spin period of \xb\ could help understand its detailed evolution. 

\subsection{Relation between Spin and Superorbital Modulations}
The spin period of \xb\ is highly variable, as shown in our time-frequency analysis.  In contrast, the superorbital modulation period is relatively stable. If the superorbital modulation were caused by the precession of a warped and tilted accretion disk \citep{OgilvieD2001}, \xb\ would belong to the group with a stable superorbital modulation similar to Her X-1, LMC X-4, and SS 433. As discussed in the previous section, the spin-down/random-walk behavior is caused by the direct wind accretion, while the disk-wind accretion is not favored. Therefore the amplitude of the superorbital modulation during the spin-down/random-walk epochs would naturally be reduced. Of the well-studied targets with a stable superorbital modulation, Her X-1 exhibited intriguing ``anomalous low states,'' in which the X-ray emission was extremely low \citep{ParmarPM1985, MiharaS1994} and the neutron star spun down \citep{ParmarOd1999}. This discovery indicates that the anomalous low state is caused by a large amount of disk warp \citep[$>90^{\circ}$, ][]{vanKerkwijkCP1998} that heavily blocked the X-ray emission from the neutron star. At the same time, this extraordinary warp results in retrograde accretion flows and causes the neutron star to spin down. However, the flux variability of \xb\ is unlike this case because the flux is anticorrelated with the spin trend, opposite to that of Her X-1. 

\citet{FarrellSO2008} proposed another possibility: the superorbital modulation might be caused by the transient mass transfer when another mass donor filled its Roche lobe near the periastron passage. If this was true, we would expect to observe a significant spin-up rate around the peak of the superorbital modulation. In this scenario, the low-mass companion orbits the neutron star with a longer period than the neutron star orbiting the massive star LS I $+$65$^{\circ}$010. Because the supergiant ($\sim 16$\,M$_{\odot}$) dominates the dynamics of this system, this scenario is possible only if the neutron star and the third companion have resonant orbits. Another possibility is that the enhanced accretion was not directly formed from the third companion but triggered by the motion of the third companion, similar to the mechanism of the superorbital modulation in 4U 1820$-$30 \citep{MazehS1979, Chou2001}. However, the stable hierarchical triple scenario requires a superorbital modulation period much longer than the observed value, making this scenario rather unlikely. \BD{Instead, the $\sim5000$-day time scale of the superorbital modulation profile variation could be caused by this effect. If this is the case, the orbital period of the third companion can be estimated by $P_{\rm{long}}=KP^2_{\rm{outer}}/P_{\rm{inner}}$ \citep{MazehS1979}, where $P_{\rm{long}}$ is the $\sim5000$-day timescale, $K$ is a constant of order unity, and $P_{\rm{inner}}$ and $P_{\rm{outer}}$ are the $11.6$\,day binary orbital period and the orbital period of the third companion, respectively. Assuming $K=1$, $P_{\textrm{outer}}$ is estimated to be $240$\,days.}

\subsection{Nature of the Orbital Profile}
Similar to the superorbital modulation, the orbital modulation period was stable but the profile was highly variable. \BD{A gradual variation of the column density during the entire orbit was observed, in which the column density gradually increased from orbital phase 0.5 to 0.9 and decreased from phase 0.1 to 0.3 \citep[see Figure 6 of][]{HallFC2000}. The sawtooth profile is therefore interpreted as the variable absorption caused by the stellar wind \citep{HallFC2000, GrundstromBG2007, FarrellSO2008}. Another interpretation is the variation in the accretion rate owing to the density gradient of the stellar wind \citep{PradhanPP2015}. The basic ideas behind them are the non-uniform distribution of the stellar wind. We found that the sawtooth profile occasionally disappeared (see e.g., Figure \ref{fold_lc_orbital}c), thereby implying the homogenizing or the ceasing of the stellar wind. At the same time, the dip was obvious}. The dip was first regarded as an eclipse; however, it is difficult to conclude that the depth of a total eclipse varies with time unless the inclination angle was at a critical value so that the pulsar was precisely obscured by the outer atmosphere of the companion. The H$\alpha$ observation suggested that the inclination angle was just below the eclipse limit and an atmospheric eclipse is possible \citep{GrundstromBG2007}. From the overlapped epoch of the ASM and BAT light curves, as well as from previous spectral studies \citep{CramptonHC1985, FarrellSO2008, PradhanPP2015}, the dip was confirmed as an absorption event rather than a total eclipse. 

Orbital-phase-dependent dipping behaviors are occasionally observed in X-ray binary systems, particularly in low-mass X-ray binaries (LMXB). The X-ray dips in LMXB systems are typically interpreted as the obscuration of X-rays by the bulge in the stream-disk interaction region \citep{WhiteS1982}. The physical interpretations of the dips in HMXB are more diverse. Cen X-3 \citep{NaikPA2011} and SMC X-1 \citep{Hu2013} are similar to the LMXBs owing to the strong evidence that these two systems exhibit accretion disks. The dips in other wind-fed HMXBs are caused by other mechanisms, including the cessation of the accretion owing to the inhomogeneous wind \citep[4U 1907+09,][]{intZandSB1997}, or the absorption by an optically thick cloud in the stellar wind \citep[Vela X-1,][]{CharlesMW1978, KreykenbohmWK2008}.  However, this type of dips usually appears intermittently and does not depend on the orbital phase. If the dips of \xb\ were not caused by the eclipse, we could consider the accretion disk scenario since this system probably had a transient accretion disk. The deep dip feature was caused by the absorption of the disk rim. We found that the dip was less significant during the spin-down/random-walk epochs (see Figures \ref{fold_lc_orbital}a and \ref{fold_lc_orbital}f). Because the direct wind accretion dominated the accretion process, the phenomena related to the disk were reduced during those epochs.

\section{Summary}\label{summary}
The source \xb\ is a unique HMXB containing one of the slowest rotating X-ray pulsars, and it shows stable orbital and superorbital modulations. With \rxte\ ASM and \swift\ BAT observations, we investigated the variation of these three modulations using the DPS and the WWZ. The evolution of the spin period showed \BD{several long-term spin-up trends with a timescale of $\sim1000$\,days and several less-significant spin-down/random-walk epochs in between. The spin-up rate indicates} that \xb\ exhibits a transient accretion disk, \BD{possibly caused by the transition from the direct wind to the disk-wind accretion.} The orbital and superorbital modulation periods are stable, but the modulation profiles are variable. The depth of the dip was highly variable, implying that the total eclipse scenario is less favored.  An obscuration by the base of the stellar wind or an obscuration by the rim of the accretion disk are possible explanations. The superorbital modulation period is also stable. However, we found that the spin-down/random-walk epochs corresponded to the increases in the X-ray flux and the decreases in the superorbital modulation amplitude. \BD{Because the accretion is dominated by the direct wind accretion during these epochs, the superorbital modulation will be less significant if it is closely related to the presence of the accretion disk.} Further monitoring in the X-ray band with a more sensitive instrument, as well as monitoring of the optical flux and radial velocity, could help to gain a better understanding of the nature of \xb.

\acknowledgments

We thank Dr. Maurizio Falanga for useful discussions \BD{and the referee for the comments that improved this paper.} This work made use of data provided by the ASM/RXTE teams at MIT and at the RXTE SOF and GOF at NASA's GSFC, \swift\ BAT data provided by the hard X-ray transient monitor \citep{KrimmHC2013}, and the MAXI data provided by RIKEN, JAXA and the MAXI team. C.-P.H.~and C.-Y.N.~are supported by a GRF grant of Hong Kong Government under HKU 17300215P.  Y.C.~is supported by the Ministry of Science and Technology (MOST) of Taiwan through the grant MOST 105-2112-M-008-012. D.C.-C.Y.~is supported by the MOST of Taiwan through the grant 105-2115-M-030-005-MY2 and an FJU project 110311041511. 

\emph{Facilities:} \rxte\ (ASM), \swift\ (BAT), MAXI



\end{document}